\documentclass[11pt]{article}
\usepackage{amsfonts,hhline}
\usepackage{amssymb,cite}

\def\noi{\noindent}
\newcommand{\foom}[1]{\protect\footnotemark[#1]}

\def\email#1#2{\footnotetext[#1]{e-mail: #2}\addtocounter{footnote}{1}}

\newcommand{\Title}[1]{\noi {\Large #1}\\[1ex]}
\newcommand{\Author}[2]{\noi{\large\bf #1}\\[2ex]\noi{\normalsize\it #2}\\}

\def\nqq{\hspace*{-2em}}

\def\nhh{\hspace*{-0.3em}}
\def\cm{\hspace*{1cm}}

\def\lal{&&\nqq {}}
\def\eq{Eq.\,}
\def\eqs{Eqs.\,}
\def\beq{\begin{equation}}
\def\eeq{\end{equation}}
\def\bear{\begin{eqnarray}}
\def\bearr{\begin{eqnarray} \lal}
\def\ear{\end{eqnarray}}
\def\earn{\nonumber \end{eqnarray}}

\def\yy{\\[5pt] {}}
\def\yyy{\\[5pt] \lal }

\def\e{{\,\rm e}}

\def\diag{\mathop{\rm diag}\nolimits}

\def\const{{\rm const}}

\def\mn{_{\mu\nu}}
\def\mN{_{\mu}^{\nu}}

\def\vac{{}_{\rm (vac)}}
\def\tot{{}_{\rm (tot)}}
\def\matt{{}_{\rm (matt)}}

\def\H{{\mathbb H}}
\def\N{{\mathbb N}}

\def\sph{spherically symmetric}
\def\ssph{static, spherically symmetric}
\def\bh{black hole}

\def\KS{Kan\-tow\-ski-Sachs}

\begin{document}

\Title {\bf Dark energy, exotic matter and properties of horizons\yy
    in black hole physics and cosmology}

\Author{K.A. Bronnikov\foom 1}
{\small Center for Gravitation and Fundamental Metrology, VNIIMS, 46
   Ozyornaya St., Moscow 119361, Russia;\\ Institute of Gravitation and
   Cosmology, PFUR, 6 Miklukho-Maklaya St., Moscow 117198, Russia}

\Author{E. Elizalde\foom 2}
{\small Consejo Superior de Investigaciones Cient\'{\i}ficas ICE/CSIC-IEEC
   Campus UAB, Facultat de Ci\`encies, Torre C5-Parell-2a pl, E-08193
   Bellaterra (Barcelona) Spain}

\Author{O.B. Zaslavskii\foom 3}
{\small Astronomical Institute of Kharkov V.N. Karazin National University,
   35 Sumskaya St., Kharkov, 61022, Ukraine}

\medskip

%%\hfill
\begin{flushright}
	{\it Dedicated to Sergei D. Odintsov\\
	     on the occasion of his 50th birthday}
\end{flushright}

%%\medskip

\begin{abstract}
  We summarize recent results on the properties of near-horizon
  metrics in different \sph\ space-times, including \KS\ cosmological models
  whose evolution begins with a horizon (the so-called Null Big Bang) and
  static metrics related to black holes. We describe the types of matter
  compatible with cosmological and \bh\ horizons. It turns out, in
  particular, that a black hole horizon can be in equilibrium with a fluid
  of disordered cosmic strings (``black holes can have curly hair''). We
  also discuss different kinds of horizons from the viewpoint of the
  behavior of tidal forces acting on an extended body and recently
  classified as ``usual'', ``naked'' and ``truly naked'' ones; in the latter
  case, tidal forces are infinite in a freely falling reference frame. It is
  shown that all truly naked horizons, as well as many of those previously
  characterized as naked and even usual ones, do not admit an extension and
  therefore must be considered as singularities. The whole analysis is
  performed locally (in a neighborhood of a candidate horizon) in a
  model-independent manner. Finally, the possible importance of some of these
  models in generating dynamic, perturbatively small vacuum fluctuation
  contributions to the cosmological constant (within a cosmological Casimir-effect
  approach to this problem) is discussed too.
\end{abstract}

%%\keywords{dark energy, Null Big Bang, black hole hair, naked black holes }
%\maketitle
\email 1 {kb20@yandex.ru}
\email 2 {elizalde@ieec.uab.es}
\email 3 {ozaslav@kharkov.ua}
%%%%%%%%%%%%%%%%%%%%%%%%%%%%%%%%%%%%%%%%%%%%%%%%%%%%%%%%%%

\section{Introduction}

  The remarkable discovery that our Universe is accelerating \cite{accel} and
  its explanation, in the framework of general relativity, in terms of the
  so-called dark energy, have posed a number of questions. The distinctive
  feature of dark energy, irrespective of its specific nature, consists in
  the violation of the standard energy conditions, including the Null Energy
  Condition (NEC). Unusual properties of this hypothetic source make us
  return to the issues which had been seemingly clarified a long time ago
  but for sources that satisfy the standard energy conditions. The
  cosmological challenge has an impact on other areas of gravitational
  physics. It concerns the existence and properties of wormholes for which
  NEC violation is necessary. In black hole physics, the necessary
  conditions of regularity include the (marginal) validity of the NEC at the
  horizon \cite{fn}, but the entire relationship between the properties of
  matter and the near-horizon geometry remains unclear. In addition, NEC
  violation can play a significant role in the possible emergence of the
  so-called truly naked black holes (TNBHs) \cite{vo,Z}, a class of objects
  in which infinite tidal accelerations in a freely falling reference frame
  is compatible with finiteness of the algebraic curvature invariants like
  the Kretschmann scalar. In cosmology, near-horizon phenomena are
  especially relevant in the context of the so-called Null Big Bang
  scenarios \cite{bd03, bd07} where the cosmological evolution itself begins
  with a horizon. Also, the possible importance of some of these
  models in generating dynamic, perturbatively small vacuum fluctuation
  contributions to the cosmological constant (within the dynamical Casimir effect
  \cite{ce} approach to this problem) will be considered, too.
  In this paper, we briefly review some recent results in this area.
  We will consider three different but related issues: Null Big Bang
  scenarios, possible black hole hair of matter characterized by macroscopic
  equations of state, and a relationship between two different
  classifications of near-horizon geometries according to their analyticity
  properties (hence extensibility beyond the horizon) and the properties of
  tidal forces acting on a freely falling body.

  In this paper, for simplicity, we restrict ourselves to \sph\ space-times,
  though extension of the results to more general geometries would surely be
  of interest. More details can be found in our works \cite{null, hair,
  horiz}.

\section{Null Big Bang and its matter sources}

  We begin our considerations with spherically symmetric cosmological models
  characterized by the general Kantowski-Sachs (KS) metric
\beq
    ds^2 =b^2 dt^2 -a^2 dx^2 -r^2 (t)d\Omega ^2, \cm
    d\Omega^2 =d\theta^2 + \sin^2 \theta d\phi^2      \label{ds-KS}
\eeq
  and supported by a source with the stress-energy tensor
\beq
        T\mN\tot = T\mN\vac + T\mN\matt,
\eeq
  where
\beq
    T\mN\vac = \diag (\rho_v,\ \rho_v,\ -p_{v\perp },\ -p_{v\perp})
            \label{SET-v}
\eeq
  describes a ``vacuum fluid'' (defined by the condition $T^0_0\vac =
  T^1_1\vac$ which guarantees invariance of $T\mN\vac$ under any Lorentz
  boosts in the distinguished $x$-direction \cite {dym92}) and
\beq
      T\mN\matt = \diag (\rho_m,\ -p_{mx},\ -p_{m\perp },\ -p_{m\perp})
            \label{SET}
\eeq
  is the contribution of matter (anisotropic fluid) taken in the most
  general form compatible with the symmetry of the metric (\ref{ds-KS}).
  We shall see that the properties of the system strongly depend on whether
  or not there is a ``vacuum'' admixture to such matter.

  In what follows, it is helpful to use the so-called quasiglobal time
  coordinate, such that $b=a^{-1}$. The coordinate defined in this way, as
  well as its counterpart in static spherically-symmetric metrics, has two
  important advantages \cite{vac1, cold}: (i) it always takes finite
  values $t=t_h$ at Killing horizons that separate static or cosmological
  regions of space-time from one another; (ii) near a horizon, the increment
  $t-t_{h}$ is a multiple (with a nonzero constant factor) of the
  corresponding increments of manifestly well-behaved Kruskal-type null
  coordinates, used for analytic continuation of the metric across the
  horizon. This condition implies the analyticity requirement for both
  metric functions $a^2(t)$ and $r^2(t)$ at $t=t_{h}$. Though, for our
  consideration, it is quite sufficient to require that these functions
  belong to class $C^2$ of smoothness.

  With this coordinate gauge, two independent combinations of Einstein's
  equations, chosen as ${0\choose 0}-{1\choose 1}$ and ${0\choose 0}$, read
  (the dot denotes $d/dt$)
\bearr
    \frac{2\ddot{r}}{r}a^2 =-8\pi (\rho _{m}+p_{mx}).  \label{01}
\yyy
    \frac{1}{r^{2}}(1+\dot{r}^2 a^2 + 2a\dot{a}r\dot{r})
                = 8\pi(\rho_m+\rho_v).       \label{00}
\ear

  Assuming the absence of interaction between matter and vacuum, the
  conservation law $\nabla_\nu T\mN =0$ should hold for each of them
  separately. Taking the component with $\mu =0$, we obtain
\beq
    \dot{\rho}_{m} + \frac{\dot a}{a}(\rho_m + p_{mx})
    +\frac{2\dot{r}}{r}(\rho_{m} + p_{m\perp })  = 0    \label{cons-m}
\eeq
  for matter and a similar equality for vacuum.
% \beq                                                        \label{cons-v}
%   \dot{\rho}_v + \frac{2\dot{r}}{r}(\rho_v + p_{v\perp }) =0.
% \eeq

  Let us assume $\rho_m \geq 0$ and consider different kinds of
  matter: ``normal'' matter that respects the NEC,
\beq
    T\mn \xi^{\mu}\xi^{\nu} \geq 0,\cm \xi_{\mu }\xi^{\mu }=0,
                                \label{NEC}
\eeq
  and ``phantom'' matter that violates it. Taking in (\ref{NEC}) the null
  vector $\xi^{\mu } = (a,\ a^{-1},\ 0,\ 0)$ we obtain the necessary
  conditions for NEC validity
\bearr
    \rho_m + p_{mx}\geq 0.       \label{NEC1}
\ear

  For normal matter, by definition, \eq (\ref{NEC1}) holds, and consequently,
  according to \eq (\ref{01}), $\ddot r \leq 0$. So we can repeat the
  argument of \cite{bd07}: let the system be expanding ($\dot r > 0$) at some
  $t_1$. Then, either $r\to 0$ at some earlier instant $t_s < t_{1}$
  (which means a curvature singularity) or the singularity is not reached,
  which can only happen due to a Killing horizon at some instant $t_h > t_s$.
  We have the following general result:

\medskip\noi
{\bf (i)} {\sl With any normal matter, regular cosmological evolution can
    only begin with a Killing horizon.}

\medskip
  Now, let us assume that there is a horizon at some $t=t_h $, so that,
  as $ t\to t_h $, $r$ remains finite while
\beq                                                        \label{hor-n}
    a^2 (t) \approx a_0 (t-t_h)^n, \cm n \in \N,
\eeq
  where $n$ is the order of the horizon. Then it immediately follows from
  (\ref{01}) and the horizon regularity requirement (which implies
  analyticity of $r(t)$ and, in particular, finiteness of $\ddot{r}$) that
\beq
    \rho_m + p_{mx}\to 0 \ \ \ \mathrm{as}\quad\ t\to t_h.   \label{hor}
\eeq
  Furthermore, we can generically assume that near the horizon the pressure
  of our matter behaves as $ p_{mx}\approx w\rho _{m}$, $w=\const$. Then
  (assuming $|p_\bot|/\rho < \infty$) \eq (\ref{cons-m}) implies the
  approximate equality
\beq
    \rho_m \approx \const\cdot a^{-(w+1)},                 \label{mhor}
\eeq
  near the horizon. This leads to the following two inferences:

\medskip\noi
  {\bf (ii)} {\sl Non-interacting normal matter cannot exist in a KS
  cosmology with a horizon; thus it can only appear there due to interaction
  with the vacuum fluid.}

\medskip\noi
  {\bf (iii)} {\sl Normal matter could only appear after a null big bang due
  to interaction with a sort of vacuum.}

\medskip
  This generalizes the conclusions made in \cite{bd07} for KS cosmologies
  with dustlike matter.

  As far as phantom matter is concerned, the presence of a Killing horizon
  is not necessary for obtaining a nonsingular cosmology. If such a horizon
  does exist, phantom matter can be present but with the restriction
  $w\leq -3$. Then, an analysis of \eqs (\ref{01})--(\ref{cons-m}) near
  the horizon shows that we can have a simple or multiple horizon with
  $\rho_v (r_h )\neq 0$ or only a simple one with $\rho _{v}(r_h )=0$. In
  such cases, a universe appearing in a Null Bang is initially contracting
  in the two spherical directions, $\dot{r}<0$.

  There is also a variant in which the Universe began its evolution
  infinitely long ago from an almost static state, which kind of evolution
  has been called ``emergent universes'' \cite{emerg}. It should be pointed
  out here that, in the KS framework, unlike isotropic cosmologies, there
  exists a variant of nonsingular evolution in which one of the scale
  factors in the metric (\ref{ds-KS}), namely, $a(t)$, vanishes as $\tau \to
  -\infty $ (where $\tau$ is the cosmological proper time, related to the
  quasiglobal time $t$ by $d\tau = \int dt/a(t)$) while the other, $r(t) $,
  remains finite in the same limit, and both timelike and null geodesics
  starting from $\tau = -\infty $ are complete. This is what can be called
  a {\it remote horizon\/} in the past, by analogy with remote horizons in
  static space-times mentioned in \cite{cold, hair}.

  We will illustrate this opportunity with two examples of such a generic
  behaviour as $\tau\to -\infty $:
\bearr
    \nqq {\bf A:}\cm   a\approx a_0  \e^{Ht} \sim 1/|\tau|,\cm
              r\approx r_0 + r_1 \e^{Ht}
\yyy
    \nqq {\bf B:}\cm   a\approx a_0 [t_0/(-t)]^q \sim |\tau|^{-q/(q+1)},\cm
          r\approx r_0 + r_1 [t_0/(-t)]^{-s},
\ear
  where $a_0 ,\ r_0 ,\ r_1,\ H,\ t_0,\ s,\ q = \const > 0$. Then, an
  analysis shows that in case A $w=-4$ and in case B $w = -3 -(s+2)/q<-3$.
  Also, in both cases, the conservation equation leads to the asymptotic
  structure of the vacuum stress tensor
\beq
    T\mN\vac = \diag (\rho_v,\ \rho_v,\ 0,\ 0).
\eeq
  Thus a combination of the Null Big Bang and emergent universe scenarios is
  possible but only under some special conditions: matter with $w\leq -3$
  and a particular structure of the vacuum stress-energy tensor in the
  remote past.

  In our reasoning, relying on the asymptotic behaviour of the density and
  pressure near the horizon, we did not assume any particular equation of
  state and even did not restrict the behaviour of the transverse pressure
  except for its regularity requirement. In this sense, our conclusions are
  model-independent. The fact that the very assumption of the existence of a
  cosmological horizon entails a number of rather general conclusions
  resembles, to some extent, the situation in black hole physics where the
  presence of the horizon greatly simplifies the description of the system
  and reduces the number of possibilities.

\section{A black hole surrounded by matter}

  In the previous section, we dealt with cosmological evolution. Now, let us
  discuss the relationship between the properties of matter and the
  near-horizon geometry in \ssph\ space-times. Such a problem arises in black
  hole physics. In real astrophysical conditions, black holes do not exist
  in empty space but are rather surrounded by some kind of matter which is
  either in equilibrium with the black hole or is falling on it. Meanwhile,
  the famous no-hair theorems (see, e.g., \cite{fn, bek} and references
  therein) are not directly applicable to such situations of evident
  astrophysical interest.

  As before, we will rely on the horizon regularity condition, the Einstein
  equations and the conservation law for matter. The manner of reasoning is
  close to that of previous section. Instead of (\ref{ds-KS}), we now have
  the metric
\beq
    ds^2 = A(u)dt^2 -\frac{du^2 }{A(u)}-r^2 (u)(d\theta^2
        +\sin^2 \theta d\phi^2),            \label{ds-u}
\eeq
  which is written using the quasiglobal radial coordinate $u$, similar to
  the quasiglobal time $t$ of Section 2 and specified by the ``gauge''
  condition $g_{00} g_{11} =-1$.  We suppose that the vacuum fluid and
  matter have the stress-energy tensor (SET) given by (\ref{SET-v}) and
  (\ref{SET}), where $p_{mx}$ (pressure of matter in the ``longitudinal''
  direction in KS cosmology) is replaced by the radial pressure $p_{mr}$.
  Note that, in \ssph\ space-times, examples of vacuum fluids are the
  cosmological constant ($p_{v\bot} = -\rho_v = \Lambda$), linear or
  nonlinear electric or magnetic fields in the radial direction ($p_{v\bot}=
  \rho_v$) and other forms which may be specified, e.g., by $\rho_v$ as a
  function of $r$ \cite{dym92, br-NED, eliz1, eliz2}.

  Two independent combinations of Einstein's equations,
  similar to (\ref{01}) and (\ref{00}), read (the prime means $d/du$)
\bearr                                                        \label{01a}
      G_0^0 - G_1^1 \equiv 2 A\frac{r''}{r}  = - 8\pi (\rho_m + p_{mr}),
\yyy
      G^1_1 \equiv                                            \label{11a}
              \frac{1}{r^2}[-1 + A'rr' + Ar'{}^2] = -8\pi(\rho_v - p_{mr}).
\ear
  We again suppose that matter and the vacuum fluid do not interact with
  each other. Then the conservation law for matter reads
\beq
    p'_r + \frac{2r'}{r} (p_{mr} - p_{m\bot})
        + \frac{A'}{2A} (\rho_m + p_{mr}) = 0.       \label{cons}
\eeq

  Now, assuming that there is a horizon at some $u=u_h $, a necessary
  condition of its regularity is that in its neighbourhood
\beq                                                        \label{hor-u}
    A(u) \approx a_0 (u-u_h)^n, \cm n \in \N,
\eeq
  where $n$ is the order of the horizon. Another regularity condition is a
  smooth (at least $C^2$) behaviour of the other metric coefficient,
  $r^2(u)$.

  One more assumption is that near the horizon the radial pressure of
  matter behaves as $ p_{mr}\approx w\rho_m$, $w=\const$. Then, on the basis
  of \eqs (\ref{01a})--(\ref{cons}) and the horizon regularity conditions,
  we can prove the following.

\medskip\noi
  {\bf Theorem 1.} {\sl A spherically symmetric black hole can be in
  equilibrium with a static matter distribution with the SET (\ref{SET})
  only if near the event horizon ($u\to u_h $, where $u$ is the quasiglobal
  radial coordinate) either (i) $w\to -1$ \textrm{(matter in this case has
  the form of a vacuum fluid)} or (ii) $ w\to -1/(1+2k)$, where $w\equiv
  p_{r}/\rho $ and $k$ is a positive integer. In case (i), the horizon can
  be of any order $n$, and $\rho (u_h)$ is nonzero. In case (ii), the
  horizon is simple, and $\rho \sim (u-u_h )^{k} $.}

  The generic case of such a non-vacuum hairy black hole is $k{=}1$,
  implying $w= -1/3$. In the case of an isotropic fluid, $p_r = p_\bot$, it
  corresponds to a distribution of disordered cosmic strings \cite{-1/3}.
  Since such strings are, in general, arbitrarily curved and may be closed,
  one can express the meaning of the theorem by the words ``non-vacuum black
  holes can have curly hair''. Recall, however, that in general our $w$
  characterizes the radial pressure, while the transverse one is only
  restricted by the condition $|p_\bot|/\rho < \infty$.

  Other values of $k$ ($k = 2,\ 3$ etc.) represent special cases obtainable
  by fine-tuning the parameter $w$.

  In the presence of vacuum matter with the SET (\ref{SET-v}), the following
  theorem holds:

\medskip\noi
  {\bf Theorem 2.} {\sl A spherically-symmetric black hole can be in
  equilibrium with a non-interacting mixture of static non-vacuum matter
  with the SET (\ref{SET}) and vacuum matter with the SET (\ref{SET-v}) only
  if, near the event horizon ($u\to u_h $), $w\equiv p_{r}/\rho \to
  -n/(n+2k)$, where  $ n\in \N$ is the order of the horizon, $n\leq k\in \N$
  and $\rho \sim (u-u_h)^k$.}

\medskip
  Thus a horizon of a static black hole can in general be surrounded by
  vacuum matter and matter with $w=-1/3$, which is true for any order of the
  horizon (i.e., including extremal and superextremal black holes) if $n=k$.
  There can also be configurations with $k>n$ and fine-tuned equations of
  state where $w=-n/(n+2k)>-1/3$. An arbitrarily small amount of other
  kinds of matter, normal or phantom, added to such a configuration, should
  break its static character by simply falling onto the horizon or maybe
  even by destroying the black hole. In other words, black holes may be
  hairy, or ``dirty'', but the possible kinds of hair are rather special in
  the near-horizon region:  normal (with $ p_r \geq 0$) or phantom hair are
  completely excluded. In an equilibrium configuration, all ``dirt'' is
  washed away from the near-horizon region, leaving there only vacuumlike or
  modestly exotic, probably ``curly'' hair.

  In particular, a static black hole cannot live inside a star of normal
  matter with nonnegative pressure unless there is an accretion region around
  the horizon or a layer of string and/or vacuum matter.

  We did not discuss the behaviour of $p_{m\bot }$ and $p_{v\bot}$. In fact,
  these quantities are inessential for our reasoning. The latter is entirely
  local, restricted to the neighborhood of the horizon, and the results,
  which involve the single parameter $w = p_r/\rho\,\big|_{\rm horizon}$, are
  in other respects model-independent. Meanwhile, a full analysis of specific
  systems would require the knowledge of the equation of state (including
  the properties of $p_{m\bot }$ and $p_{v\bot}$) and conditions on the
  metric in the whole space (e.g., the asymptotic flatness condition). Such
  an analysis depends on the model in an essential way and is beyond the
  scope of this paper. One can add that the equations of state well-behaved
  near the horizon are often incompatible with reasonable conditions at
  infinity (see, e.g., the example of an exact solution with string fluid in
  \cite{hair}); it simply means that such matter does not extend to infinity
  and can only occupy a finite region around the horizon.

  Our inferences are quite general and hold for all kinds of hair: for
  instance, in all known examples of black holes with scalar fields (see,
  e.g., \cite {pha1} and references therein), the SETs near the horizon must
  satisfy the above conditions, which may be directly verified.

  Also, our approach is relevant to semiclassical black holes in equilibrium
  with their Hawking radiation (the Hartle-Hawking state), whose SET
  essentially differs from that of a perfect fluid. Since the density of
  quantum fields is, in general, nonzero at the horizon (see Sec.\,11 of
  the textbook \cite{fn} for details), the regularity condition (\ref{hor})
  (with $t$ replaced by $r$) tells us that such quantum radiation should
  behave near the horizon like a vacuum fluid. Our results show that a black
  hole can be in equilibrium with a mixture of Hawking radiation and some
  kinds of classical matter with $-1<w<0$ (including the important case of a
  Pascal perfect fluid with $p_r = p_{\bot }$).  Possible effects of this
  circumstance for semiclassical black holes need a further study. Moreover,
  large enough black holes, for which the Hawking radiation may be
  neglected, can be in equilibrium with classical matter alone, also
  including the case of a perfect fluid.

  It would be of interest to generalize our results to nonspherical and
  rotating distributions of matter.

\section{Truly naked horizons and their sources}

  In the previous two sections, the restriction on possible matter sources
  supporting geometries with Killing horizons essentially relied on the
  horizon regularity condition, which essentially meant analyticity.
  Meanwhile, the notion of regularity is by itself not as obvious as one
  could think. In particular, it turns out that there exist such horizons
  that all scalars composed algebraically from the components of the
  curvature tensor are finite there but some separate curvature components
  (responsible for the transverse tidal forces) enormously grow when
  approaching the horizon \cite{cold, h-ross, horiz, tr, qbh}. From a
  mathematical viewpoint, such cases represents interesting examples of
  so-called nonscalar singularities \cite{e-sch77}. This makes especially
  important a careful analysis of the metric near the surfaces which can be
  called candidate horizons and a comparison between the properties of
  tidal forces on such surfaces and the conditions under which the metric
  can be extended beyond them.

  Let the metric in the Schwarzschild-like coordinates be written as
\beq
    ds^2 = \e^{2\gamma }dt^2 - \e^{2\alpha }dr^2 -r^2 d\Omega^2.
                            \label{ds-r}
\eeq
  Let us assume that near a candidate horizon $\H:\ r=r_h$ (where, by
  definition, $\e^\gamma \to 0$) the metric coefficients behave as follows:
\beq
    \e^{2\gamma}\sim (r-r_h)^q,\cm \e^{2\alpha}\sim (r-r_h)^p,
\eeq
  with $p>0$ and $q>0$. As follows from the geodesic deviation equations,
  the tidal forces experienced by bodies in the gravitational field are
  conveniently characterized by the combination of components of the
  curvature tensor $Z:=R^{12}{}_{12}-R^{02}{}_{02}$ in the static reference
  frame and by $\bar{Z} = Z\e^{-2\gamma}$ in a freely falling reference
  frame near $\H$. These quantities have been used in \cite{Z} to distinguish
  usual ($Z = 0 = \bar{Z}$,), naked ($Z=0$, $\bar{Z}\neq 0$ is finite)
  and truly naked ($Z=0$, $\bar{Z} = \infty$) horizons. In all cases we
  consider surfaces $\H$ at which all algebraic curvature invariants are
  finite, and this is so under the condition
\beq
    p\geq 2\ \ \mathrm{or} \ \ \ 2 > p\geq 1,  \ \ p+q=2.  \label{reg-r}
\eeq

  The comparison is carried out by rewriting the metric (\ref{ds-r}) in
  terms of the quasiglobal coordinate $u$ [\eq (\ref{ds-u})] and imposing
  the requirement that the metric coefficients $A(u)$ and $r^2(u)$ should be
  analytic at $\H$, where $A(u) = \e^{2\gamma (r)} =0$. In particular, we
  obtain the condition
\beq
      q(n-2) = n (p-2).                                    \label{pq1}
\eeq
  which selects a sequence of lines in the ($p,q$) plane, intersecting at
  the point $(-2,0)$.

  The results of such a comparison are presented in Table 1 \cite{horiz}.

\begin{table} %%
  \caption{Horizon types according to the properties of tidal forces and
           regularity (extensibility) of the metric; $n \in \N$ is the order
       of the horizon.}

\medskip
\centering
\renewcommand{\arraystretch}{1.15}
\tabcolsep 4pt
\begin{tabular}{|c|l|l|l|}
\hline
  No.\nhh &  $p,\ q$    & type by tidal forces & regularity \\
\hhline{|=|=|=|=|}
  1   & $p = q = 1$ & usual or naked  & regular, $n=1$  \\
\hline
  2   & $1 < p < 3/2$, $q=2-p$
            & truly naked        & singular         \\
\hline
  3   & $p=3/2,\ q= 1/2$
            & naked              & regular, $n=1$    \\
\hline
  4   & $ 3/2 < p < 2$, $q=2-p$
            & usual              & regular, $n=1$    \\
\hline
  5   & $p \geq 2,\ q > p$
            & truly naked        & singular         \\
\hline
  6   & $p \geq 2,\ q=p$
            & usual or naked     &
                        regular if $p=q=n$, otherwise singular \\
\hline
  7   & $p \geq 2,\ p-1 < q < p$
            & truly naked        & singular         \\
\hline
  8   & $p \geq 2,\ q = p-1$
            & naked              &
                    regular if $p=1+n/2$, otherwise singular \\
\hline
  9   & $p \geq 2,\ p-2 < q < p-1$
            & usual              &
              regular if (\ref{pq1}) holds, $n\in \N$, otherwise singular\\
\hline
  10  & $p \geq 2,\ q \leq p-2$
            & usual              & remote horizon   \\
\hline
\end{tabular}
\end{table}

  We see from the table that all truly naked horizons are, in fact,
  singularities since the metric cannot be extended beyond them; moreover,
  some naked and even usual horizons turn out to be singular.

  As to possible source of gravity leading to different types of horizons,
  the situation turns out to be the following. If we consider arbitrary
  one-component matter with $p_r/\rho = w =\const \ne -1$ (at least near the
  surface $\H$), the only possible solutions correspond to a simple horizon,
  such that $A(u)\sim u-u_h $, and, provided $w=-1/(1+2k)$ where $k$ is a
  positive integer, we obtain regular solutions in full agreement with
  Section 3. Solutions with truly naked horizons are not obtained.

  If, however, we consider a mixture of the two kinds of matter described by
  (\ref{SET}) and (\ref{SET-v}), there appear solutions containing matter
  with $0 > w >-1$ and $\rho \sim A^{-(w+1)/(2w)}$. Furthermore, if we turn
  to the curvature coordinates, we obtain, for $p\neq q$, the relation
  $w=-q/(q+2p-2)$, and it appears that $\rho >0$ for $q>p$ and $\rho <0$ for
  $q<p $. Thus, any $p$ and $q$ satisfying the condition (\ref{reg-r}) are
  admissible, except for those with $p=q$. In the latter case, solutions can
  also exist, with $w$ satisfying the requirement $w>-p/(3p-2)$.
  All kinds of solutions mentioned in Table I are possible, and the values
  of $w$ cover the whole range from 0 to $-1$. This is related to the
  underdetermined nature of the system since the function $\rho\vac(u)$
  remains arbitrary.

  If we put $\rho _{v}=\Lambda /(8\pi) = \const$, thus specifying the vacuum
  as a cosmological constant, the Einstein equations relate the exponents
  $p$ and $q$ characterizing the metric to the matter parameter $w$. Namely,
  we have either (i) $p=q=1$ (a simple regular horizon) and $w=-1/(1+2k)$
  (as described in Section 3 and \cite{hair}) or (ii) $p=2$, $q=-2w/(w+1),$
  $w\neq -1/2$.  The parameter $w$ can take any value in the range $(-1,\ 0)$
  except $-1/2$.

\section{Conclusion}

  In a model-independent way, we have etablished the correspondence between
  the equation of state (in terms of the parameter $w= p_r/\rho$) and the
  type of horizon both for cosmological scenarios and for black holes. We
  found the interval of $w$ for which regular or (truly) naked horizons
  occur. Certain discrete values of $w$ characterize possible ``hair''
  around a regular black hole horizon. Thus we have used a unified
  approach to so seemingly different physical objects and phenomena as Null
  Big Bang, the hair properties of black holes and (truly) naked black
  holes.

  This consideration, along with \cite{bd07, null}, suggests an interesting
  type of cosmological scenarios, with such stages as (i) a static or
  stationary core, (ii) a de Sitter-like horizon, (iii) particle creation
  and isotropization, (iv) a hot stage and further on according to the
  Standard Model.

  In the context of the early Universe, in addition to particle creation,
  it would be of interest to take into account one more quantum phenomenon,
  the ordinary \cite{ce1} and the dynamical \cite{ce2} Casimir effects
  related to the nontrivial topology of KS
  models, e.g., in the manner of Refs.\,\cite{NO99,ocas1}, and its possible
  influence on the structure of singularities like those discussed in this
  paper. It has been argued that Casimir considerations can play no role in
  trying to solve the problem of the cosmological constant, in its hard form.
  This seems actually to be true, but provided a drastic suppression of the
  main contributions to the same for some particular topology could be
  proven to happen in some model, then additional, sort of perturbative
  contributions coming from some adjustments in the topology or the evolution
  of our universe could provide a clue to solve the issue of its value
  being so small. It is in this context that Casimir-like calculations as
  mentioned could be of importance. At the very least, proving that these
  additional contributions are of the same order of magnitude as the observed
  value of the universe acceleration is already a first step, that has been
  undertaken in some specific cases \cite{cccas1}.

  It would also be of interest to relate the origin of singularities in KS
  cosmology subject to quantum effects with their effective 2D description,
  in the manner of \cite{nooo99}, where quantum-corrected KS cosmologies were
  investigated. The effective 2D description makes the presentation
  qualitatively easier and may reveal a fundamental structure behind
  singularities, related to quantum effects.

  As to singular horizons in KS cosmology as discussed in this paper (at
  least simple ones), they can be considered as examples of the so-called
  finite-time singularities. Four types of such singularities are known and
  classified for isotropic FRW models in \cite{NOTs05}. Among them, the Big
  Rip (or type I singularity) is the most well-known and is widely discussed
  in connection with different models of dark energy. It is clear that in KS
  models, where we have two scale factors, such singularities may occur and
  their properties should be more diverse, and the corresponding
  classification should be naturally extended as compared with the
  one-scale-factor FRW cosmology. Such an extended classification, which
  should also apply to static, spherically symmetric analogues of KS
  cosmologies as well as to other, more complicated anisotropic cosmologies,
  is of significant interest. We hope to present a detailed description of
  such singularities in our future publications.

  It is a pleasure for us to dedicate this paper to Sergei
  Odintsov on the occasion of his 50th birthday.

\subsection*{Acknowledgement}

  E.E. has been supported in part by MEC (Spain), project FIS2006-02842,
  and by AGAUR (Gene\-ra\-litat de Catalunya), contract 2005SGR-00790.
  K.B. has been supported by the Russian Basic Research Foundation grant  
  No. 07-02-13624-ofi-ts and by a grant of People Friendship University 
  of Russia (NPK MU).
  The work of O.Z. was supported by European Science Foundation, Short
  Visits and Exchange Programme, grant \# 2536.

\end{document}